\documentstyle{mn}

\newif\ifAMStwofonts

\ifoldfss
  \ifCUPmtlplainloaded \else
    \NewTextAlphabet{textbfit} {cmbxti10} {}
    \NewTextAlphabet{textbfss} {cmssbx10} {}
    \NewMathAlphabet{mathbfit} {cmbxti10} {} 
    \NewMathAlphabet{mathbfss} {cmssbx10} {} 
  \fi
  \ifAMStwofonts
    \ifCUPmtlplainloaded \else
      \NewSymbolFont{upmath} {eurm10}
      \NewSymbolFont{AMSa} {msam10}
      \NewMathSymbol{\upi}     {0}{upmath}{19}
      \NewMathSymbol{\umu}     {0}{upmath}{16}
      \NewMathSymbol{\upartial}{0}{upmath}{40}
      \NewMathSymbol{\leqslant}{3}{AMSa}{36}
      \NewMathSymbol{\geqslant}{3}{AMSa}{3E}

       \let\le=\leqslant
       
    \fi
  \fi
\fi 

\ifnfssone
  \newmathalphabet{\mathit}
  \addtoversion{normal}{\mathit}{cmr}{m}{it}
  \addtoversion{bold}{\mathit}{cmr}{bx}{it}
  \newmathalphabet{\mathbfit} 
  \addtoversion{normal}{\mathbfit}{cmr}{bx}{it}
  \addtoversion{bold}{\mathbfit}{cmr}{bx}{it}
  \newmathalphabet{\mathbfss} 
  \addtoversion{normal}{\mathbfss}{cmss}{bx}{n}
  \addtoversion{bold}{\mathbfss}{cmss}{bx}{n}
  \ifAMStwofonts
    \ifCUPmtlplainloaded \else
      \UseAMStwoboldmath
      \makeatletter
      \new@mathgroup\upmath@group
      \define@mathgroup\mv@normal\upmath@group{eur}{m}{n}
      \define@mathgroup\mv@bold\upmath@group{eur}{b}{n}
      \edef\UPM{\hexnumber\upmath@group}
      \new@mathgroup\amsa@group
      \define@mathgroup\mv@normal\amsa@group{msa}{m}{n}
      \define@mathgroup\mv@bold\amsa@group{msa}{m}{n}
      \edef\AMSa{\hexnumber\amsa@group}
      \makeatother
      \mathchardef\upi="0\UPM19
      \mathchardef\umu="0\UPM16
      \mathchardef\upartial="0\UPM40
      \mathchardef\leqslant="3\AMSa36
      \mathchardef\geqslant="3\AMSa3E

       \let\le=\leqslant

    \fi
  \fi
\fi 

\ifnfsstwo
  \DeclareMathAlphabet{\mathbfit}{OT1}{cmr}{bx}{it}
  \SetMathAlphabet\mathbfit{bold}{OT1}{cmr}{bx}{it}
  \DeclareMathAlphabet{\mathbfss}{OT1}{cmss}{bx}{n}
  \SetMathAlphabet\mathbfss{bold}{OT1}{cmss}{bx}{n}
  \ifAMStwofonts
    \ifCUPmtlplainloaded \else
      \DeclareSymbolFont{UPM}{U}{eur}{m}{n}
      \SetSymbolFont{UPM}{bold}{U}{eur}{b}{n}
      \DeclareSymbolFont{AMSa}{U}{msa}{m}{n}
      \DeclareMathSymbol{\upi}{0}{UPM}{"19}
      \DeclareMathSymbol{\umu}{0}{UPM}{"16}
      \DeclareMathSymbol{\upartial}{0}{UPM}{"40}
      \DeclareMathSymbol{\leqslant}{3}{AMSa}{"36}
      \DeclareMathSymbol{\geqslant}{3}{AMSa}{"3E}

       \let\le=\leqslant

    \fi
  \fi
\fi 

\ifCUPmtlplainloaded \else
  \ifAMStwofonts \else 
    \def\upi{\pi}
    \def\umu{\mu}
    \def\upartial{\partial}
  \fi
\fi

\title{Production of neutrons, neutrinos and gamma-rays by a very fast 
pulsar in the Galactic Center region} 
\author[W. Bednarek]
       {W. Bednarek \\
        Department of Experimental Physics, University of \L \'od\'z,
        ul. Pomorska 149/153, 90-236 \L \'od\'z, Poland}
\date{Accepted 
      Received ;
      in original form }
\pubyear{}

\begin{document}

\maketitle

\begin{abstract}
We consider the possibility that the excess of cosmic rays near 
$\sim 10^{18}$ eV, reported by the AGASA and SUGAR groups from the 
direction of the Galactic Center, is caused by a young, very fast 
pulsar in the high density medium. The pulsar accelerates iron nuclei
to energies $\sim 10^{20}$ eV, as postulated by the Galactic 
models for the origin of the highest energy cosmic rays. The iron nuclei,
after about 1 yr since pulsar formation, leave the supernova
envelope without energy losses and diffuse through the dense 
central region of the Galaxy. Some of them collide with the background 
matter creating neutrons (from desintegration of Fe), neutrinos and
gamma-rays (in inelastic collisions).
We suggest that neutrons produced at a specific time after the pulsar 
formation
are responsible for the observed excess of cosmic rays at $\sim 10^{18}$ eV. 
From normalization of the calculated neutron flux to the one observed 
in the cosmic ray excess, we predict the neutrino and gamma-ray fluxes.
It has been found that the 1 km$^2$ neutrino detector of the IceCube type
should detect from a few up to several events per year from the Galactic 
Center, depending on the parameters of the considered model. Also future
systems of Cherenkov telescopes (CANGAROO III, HESS, VERITAS) should be able to
observe 1 - 10 TeV $\gamma$-rays from the Galactic Center if the pulsar was
created  inside a huge molecular cloud about $3-10\times 10^3$ yrs ago.

\end{abstract}


\section{Introduction}

Recently the AGASA collaboration has reported the existence of an extended
excess of cosmic rays (CRs) over a narrow energy range $10^{17.9} - 10^{18.3}$ eV
from directions close to the Galactic Center (GC) and the Cygnus region with
the significance of 4.5$\sigma$ and $3.9\sigma$, respectively (Hayashida et
al.~1999). The GC excess was confirmed in the analysis of the SUGAR 
data (Bellido et al.~2001), in which case the observed signal in the energy
range $10^{17.9} - 10^{18.5}$ eV is consistent with that from a point-like
source of neutral particles, with an estimated flux of $(9 \pm 3)\times
10^{-14}$ m$^{-2}$ s$^{-1}$, offset from the true location of the GC by about
$7.5^{\rm o}$. Hayashida et al.~(1999) suggested that such an anisotropy of
CRs  in a narrow energy range can be explained naturally by
neutrons due to the fact that particles with such energies are able to reach
the Earth before decaying from distance of the GC. Numerical simulations 
of charged particles (protons) propagation in reasonable models of the
Galactic magnetic field give rise to the more extended source which may 
even be significantly shifted in the sky from the original direction towards
the source (e.g. Clay~2000,  Bednarek, Giller \& Zieli\'nska~2001). However,
the results of charged hadrons propagation strongly depend on the details
of the magnetic field model which is not well known at present.

The production of neutrons in discrete galactic sources  
was already discussed some time ago by Jones~(1990), Sommers \& Elbert~(1990)
and Bednarek~(1992) in relation to reports of the excess of EeV particles from
the direction of Cygnus X-3 (Cassiday et al.~1989, Teshima et al.~1990).
It has recently been suggested that in the GC region
neutrons might be produced in collisions of hadrons with matter 
(Takahashi \& Nagataki~2001). Hadrons
can be accelerated by a massive black hole associated with the Sgr A$^*$
(Levinson \& Boldt~2000), or by the shock waves of supernovae which explode
into their own stellar winds (Rhode, Enslin \& Biermann~1998). 

The Galactic Center region (inner $\sim 50$ pc) is rich in many massive
stellar clusters with a few to more than 100 OB stars (Morris \& Serabyn~1996, 
Blum et al.~2001). These stars should soon explode as supernovae. In fact
recent multiple supernova explosions in the GC region ($10^3$ supernovae in
the past $10^5$ years) are suggested by the observations of the diffuse hot
plasma emitting X-rays (Yamauchi et al.~1990). For example one remnant of
such a young supernova with the age of $\sim 80$ years (G0.570-0.018) has 
recently been reported by Senda et al.~(2001). Since it is expected that 
pulsars are formed in explosions of such massive stars, we can expect that the
GC region should contain some young pulsars, a number of them being
$10^2 - 10^3$ yrs old.  Motivated by these observational results we may assume
that at least one of these young pulsars, formed in a supernova explosion of
the type Ib/c, has parameters which allow the acceleration of iron
nuclei to energies $\sim 10^{20}$ eV, as postulated by some models for the
Galactic origin of the highest energy cosmic rays  (Blasi, Epstein \&
Olinto~2000; De Goubeia Dal Pino \& Lazarian~2000).  We suggest that 
neutrons from desintegration of iron nuclei, which are  accelerated by such
energetic pulsar in the GC region, can be responsible for the  observed excess
of the cosmic rays with energies $\sim 10^{18}$ eV. Note that the iron nuclei
with energies $\sim 10^{20}$ eV and neutrons with energies $\sim 2\times
10^{18}$ eV have the same Lorentz factors.  
In order to test this hypothesis we predict the neutrino and
$\gamma$-ray fluxes accompaning the process of neutron injection by iron
nuclei. 

\section{A pulsar inside a molecular cloud}

We investigate the scenario in which a very young pulsar is formed 
in a core collapse of the type Ib/c supernova immersed within
a huge molecular cloud (or high density medium), characteristic of the GC
region. Such supernovae are probably projenitors of neutron stars with 
extreme parameters (period, magnetic field surface), which allow them to
accelerate iron nuclei to the highest energies observed in the CRs.
In about a year after supernova explosion, the pulsar 
is surrounded by such a dense expanding envelope that hadrons, accelerated in
the pulsar magnetosphere or the pulsar wind zone, are not injected into the
surrounding. These hadrons lose energy on multiple inelastic collisions 
producing high energy neutrinos (Protheroe, Bednarek \&
Luo~1998, Bednarek~2001, Beall \& Bednarek~2001). In later stages the
particles can escape into the surrounding and diffuse in the magnetic field of
the cloud, suffering collisions with the matter from time to time. In our
further considerations we discuss, as an example, two media typical of
the GC region in which the pulsar may be immersed. The first one is a huge
molecular cloud with  the radius $R_{\rm c} = 10$ pc, the density $n_{\rm c} =
10^3$ cm$^{-3}$, and  the magnetic field $B_{\rm c} = 10^{-4}$ G (
the total mass
$\sim 10^5 M_\odot$), and the second one is an extended high density
region inside the GC with $R_{\rm c} = 50$ pc, $n_{\rm c} = 10^2$ cm$^{-3}$,
and $B_{\rm c} = 3\times 10^{-5}$ G (the total mass $\sim 10^6 M_\odot$). The
clouds with such parameters are typical of the GC region and of the Nuclear
Bulge, which is a narrow layer of the interstellar matter in the central
$\sim 600$ pc of the Galaxy with the height of $\sim 50$ pc.

\subsection{Acceleration of nuclei by a pulsar}

Following the recent works of Blasi, Epstein \& Olinto (2000), Beall \&
Bednarek~(2001), and Bednarek~(2001) we assume that pulsars can accelerate
iron nuclei in its wind zone (the mechanism called magnetic slingshot, Gunn
\& Ostriker~1969).
The maximum energies which the nuclei can reach in this model, are determined
by the magnetic field energy per particle in the pulsar wind zone and depend on
the pulsar parameters as,      
\begin{eqnarray}
E_{\rm Fe} = {{B^2(r_{\rm LC})}\over{8\pi n_{\rm GJ}(r_{\rm LC})}}
\approx 1.8\times 10^{11} B_{12} P_{\rm ms}^{-2}~{\rm GeV}, 
\label{eq1}
\end{eqnarray}
\noindent
where $P = 10^{-3}P_{\rm ms}$ s is the pulsar period, $B = 10^{12}
B_{12}$ G is the pulsar's surface magnetic field, $n_{\rm GJ} = B(r_{\rm
LC})/(2 e c P)\approx 3.3\times 10^{11} B_{12} P_{\rm ms}^{-4}$ cm$^{_3}$
is Goldreich \& Julian (1969) density, $r_{\rm LC} = cP/2\pi 
\approx 4.77\times 10^6 P_{\rm ms}$ cm, and $c$ is the velocity of light.
According to the slingshot
mechanism, the acceleration of nuclei occurs very fast so that they do not lose
energy during this stage. During the first $\sim 1$ yr after explosion
these nuclei interact with the radiation field and matter of the supernova
envelope (Beall \& Bednarek~2001).  Only when the envelope becomes transparent,
can the iron nuclei be injected into the surrounding of the parent pulsar's
supernova. We obtain the spectrum of iron nuclei injected by the pulsar
following the derivation by  Beall \& Bednarek~(2001). Their calculation
is based on the general prescription given by Blasi, Epstein \& Olinto~(2000)
in which the number of nuclei accelerated to energies $E$ scales as a part
$\xi$ of the Goldreich \& Julian (1969) density at the light cylinder radius. 
Beall \& Bednarek~(2001) modify this approach  by noting that the  spectrum
injected by the pulsar at a fixed time 't' should be non-monoenergetic due
to the fact that the magnetic energy density, responsible for the acceleration
of  particles, changes along the pulsar's light cylinder height.  
As a result, a pulsar with specific parameters injects nuclei with the 
spectrum which can be below $E_{\rm Fe}$ described by (see
details in Beall \& Bednarek~2001),  %
\begin{equation}
\begin{array}{l}
{{dN}\over{dEdt}} = {{2\pi c \xi r_{\rm LC}^2 n_{\rm GJ}
(E_{\rm Fe} E^2)^{-1/3}}\over{3\left[(E_{\rm Fe}/E)^{2/3} -
1\right]^{1/2}}} \\
~~~~~~\cong  {{3\times 10^{30}\xi (B_{12} P_{\rm
ms}^{-2}E^{-1})^{2/3}} \over{\left[(E_{\rm Fe}/E)^{2/3} - 1\right]^{1/2}}}
~{\rm {{Fe}\over{s~GeV}}}.     
\label{eq2}
\end{array}
\end{equation} 

We estimated that these nuclei can escape through the supernova envelope 
after $\sim 1$ yr after the supernova explosion for typical parameters of the 
supernova, i.e the mass of the envelope in the case of type Ib/c supernovae
$M_{\rm env} = 3M_\odot$, and the expansion velocity of the envelope at the
inner radius is $v_{\rm env} = 3\times 10^8$ cm s$^{-1}$ (Beall
\& Bednarek~2001). The iron nuclei diffuse in the magnetic field of the high
density medium in the GC region, i.e. huge molecular clouds. Some of them
interact producing neutrons, neutrinos, and $\gamma$-rays.  

In order to obtain the equilibrium spectrum of iron nuclei inside the cloud,
we have to integrate over the activity time of the pulsar 
since its parameters evolve in time due to the pulsar's energy losses.
If we assume that the pulsar loses energy only on
electromagnetic waves, then its period changes according to the formula
$P_{\rm ms}^2(t_{\rm obs}) = 1.04\times 10^{-9} t_{\rm obs}B_{12}^2 + P_{\rm
0,ms}^2$, where  $P_{\rm 0,ms}$ $P_{\rm ms}(t_{\rm obs})$ are the initial 
and present periods of the pulsar. The equilibrium spectrum of iron nuclei 
at a specific observation time, $t_{\rm obs}$, is calculated from 
\begin{eqnarray}
{{dN}\over{dE}} = \int_{t_{\rm 0}}^{t_{\rm obs}} 
{{dN}\over{dE dt}} K e^{-c(t_{\rm obs} - t)/\lambda} dt,
\label{eq3}
\end{eqnarray}
\noindent
where $t_{\rm 0} =1$ yr, $K$ gives the part of
nuclei produced at the time 't' which do not escape from the cloud due to
the diffusion  and are still present inside the cloud at the time $t_{\rm
obs}$. $\lambda$ is the mean free path for collision of the iron nuclei with
the matter of the cloud. Note that for the parameters considered in this paper
the  value $c(t_{\rm obs} - t)/\lambda$, in Eq.~\ref{eq3}, is always less than
one. Therefore, a significant part of iron nuclei with energies $\sim 10^{20}$
eV can escape without interaction as postulated by the model of Blasi, Epstein
\& Olinto~(2000). The value of $K$
is estimated from $K = (R_{\rm c}/D_{\rm dif})^3$, where 
$D_{\rm dif} = (r_{\rm L}ct/3)^{1/2}$ is the
diffusion  distance of iron nuclei in the magnetic field of the cloud, and
$r_{\rm L}$ is the Larmor radius of the iron nuclei with energy $E$. For the
case,  $D_{\rm dif} \le R_{\rm c}$, we take $K = 1$. We assume that the giant
molecular cloud does not expand. So then the nuclei do not suffer adiabatic
energy losses.

\subsection{Interaction of nuclei inside the cloud}

The part of iron nuclei confined within the molecular cloud, interact with
a relatively dense medium suffering desintegrations and pion energy losses.
The pions decay into neutrinos and $\gamma$-rays.
Applying the equilibrium spectrum of iron nuclei (see Eq.~\ref{eq3}), we
calculate the differential spectra of neutrons (from desintegrations of the
iron nuclei), muon neutrinos, and $\gamma$-rays (from inelastic collisions of
iron). In these calculations we assume that: (1) in a single interaction, the
iron nuclei desintegrate into two nuclei and the rest of   neutrons is
released with the Lorentz factors comparable to  the Lorentz factor of the
primary nuclei; (2) pions, produced in interactions of iron nuclei with the 
matter of the cloud, have Lorentz factors comparable to their parent nuclei.
They are produced with  multiplicities given in Orth \& Buffington~(1976).

\begin{figure}
  \vspace{6.truecm}
\includegraphics{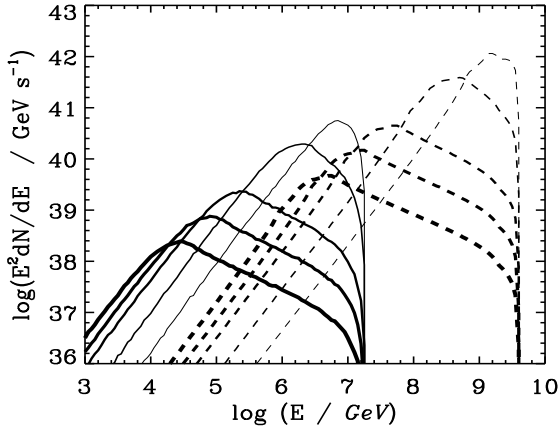}
  \caption{The differential spectra of neutrons (dashed curves)
and $\gamma$-rays (full curves) produced in the
interactions of iron nuclei with the matter of a molecular cloud with the
radius $R_{\rm c} = 10$ pc, density $n_{\rm c} = 10^3$ cm$^{-3}$, and the
magnetic field $B_{\rm c} = 10^{-4}$ G, at the time after the pulsar
formation: t = 10, $10^2$, $10^3$, $3\times 10^3$, and $10^4$ yr (from the
thinnest to the thickest curve). The initial parameters of the pulsar are 
$P_{\rm ms} = 2$ and $B_{12} = 6$.} 
\label{fig1}
\end{figure}

The computations of differential neutron, neutrino, and $\gamma$-ray
spectra have been done for the pulsar with the surface magnetic field 
$B_{12} = 6$ (typical of the observed radio pulsars) and the initial 
period $P_{\rm ms} = 2$. Such a pulsar is able to accelerate nuclei to 
such high energies (see Eq.~\ref{eq1}) that neutrons from their
desintegration fulfil the observational constraint
put by the excess of cosmic ray particles at $10^{7.9} - 10^{8.5}$ GeV
(Hayashida et al.~1999, Bellido et al. 2001).
The pulsar born with such parameters slows down due to the dipole
energy losses to 11 ms after $10^2$ yr, 34 ms ($10^3$ yr), 60 ms 
($3\times 10^3$ yr), and to 106 ms  ($10^4$ yr).
So then it is still fast enough to produce pulsaed $\gamma$-rays
(Crab and Vela type pulsars).

As an example in Fig.~\ref{fig1} we show the spectra of neutrons 
(dashed curves) and $\gamma$-rays (full curves) produced by the iron nuclei
inside a cloud with the radius $R_{\rm c} = 10$ pc, the density $n_{\rm c} =
10^3$ cm$^{-3}$ (the cloud mass $10^5 M_\odot$), and the magnetic field  
$B_{\rm c} = 10^{-4}$ G at different times after the pulsar formation 
$t_{\rm obs} = 10, 10^2, 10^3, 3\times 10^3$, and  $10^4$ yrs.
As expected, the spectra of neutrons extend up to a few $10^{18}$ eV. 
However, their intensities at these highest energies significantly drop with
time after the pulsar formation due to the escape of iron nuclei from the 
cloud for the time $t_{\rm obs} > 10$ yr, which is caused by their diffusion in
the magnetic field. Due to the large multiplicities of pion
production at these high energies, the maximum energies of neutrinos and
$\gamma$-rays are about 2 orders of magnitude lower than for neutrons. 
Their spectra at the higiest energies also drop with time due to the escape 
of nuclei from the cloud.

\begin{figure}
  \vspace{12truecm}
\includegraphics{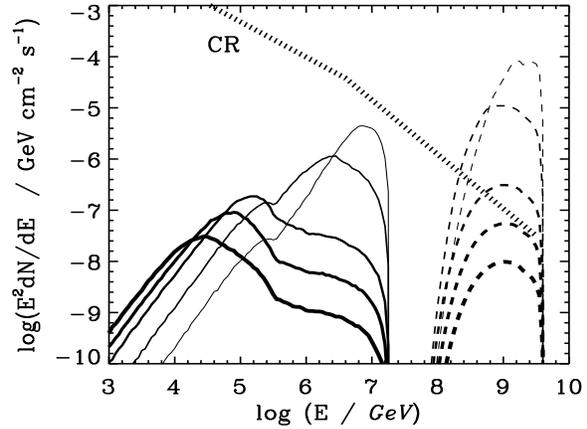}
\includegraphics{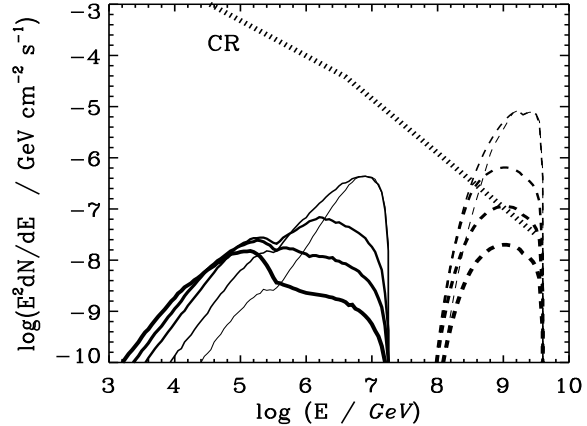}
  \caption{The differential spectra of neutrons (dashed curves) and
$\gamma$-rays (full curves) observed from the Galactic Center on Earth
at the time $10, 10^2, 10^3, 3\times 10^3$, and $10^4$ yrs (from the
thinnest to the thickest curve) after the formation of the pulsar. The initial
parameters of the pulsar are as in Fig.~\ref{fig1}.
The thick dotted curve shows schematically the observed cosmic ray 
spectrum (see e.g. Boratav \& Watson~2000) within the 20$^{\rm o}$ circle.}  
\label{fig2}
\end{figure}
\section{Neutrons, neutrinos, and gamma-rays on Earth}

The large distance to the GC, $D_{\rm GC}\approx 8.5$ kpc, has influence on the
expected fluxes of neutrons and $\gamma$-rays
measured on Earth, due to the decay of unstable neutrons and the 
absorption of $\gamma$-rays in the microwave background radiation (MBR). 
The mean free path for neutrons depends only on their energy,
$\lambda_{\rm N} = c\tau_{\rm N}\gamma_{\rm N}\approx  9.8 E_{\rm N, EeV}$ 
kpc, where a neutron's lifetime is $\tau_{\rm N} = 918$ s, its energy 
$E_{\rm N} = 9.38\times 10^{-10}\gamma_{\rm N}$ in EeV, and  $\gamma_{\rm N}$
is the Lorentz factor. The mean free path for the absorption of photons in MBR,
$\lambda_{\gamma\gamma}$, was computed in several papers just after the
discovery of MBR (e.g. Gould \& Schreder~1966). $\lambda_{\gamma\gamma}$ 
becomes comparable to the distance of the GC at energies of $\sim 1$ PeV.
Therefore the fluxes of neutrons and  $\gamma$-rays observed on Earth are
reduced by the  factor  ${\rm exp}(-D_{\rm GC}/\lambda_{\rm \gamma\gamma; N})$,
where $\lambda_{\rm \gamma\gamma; N} = \lambda_{\rm \gamma\gamma}$ or 
$\lambda_{\rm N}$ respectively. In
Fig.~\ref{fig2}, we show the fluxes of neutrons and $\gamma$-rays observed 
on Earth at different times after the pulsar formation for the case of the
pulsar with the previously mentioned initial parameters and two different 
sets of parameters describing the medium in the GC region mentioned in
section~2: the case of the molecular cloud (the upper figure), and the case of
a more extended and lower density region around the GC (the lower figure). 
The comparison of Figs.~\ref{fig1}, and~\ref{fig2} allows us to conclude 
that the absorption effects of $\gamma$-rays become important 
in the PeV energy range, and that neutrons with energies
below  $\sim 10^8$ GeV are not able to reach the Earth.
Since the propagation of neutrinos is not influenced by
any  process, their fluxes can be simply obtained for the known distance to
the GC. In Fig.~\ref{fig2} we mark also the observed spectrum of cosmic rays
(CR)  below $3\times 10^{18}$ eV within the $20^{\rm o}$ circle (the analysis
box of the AGASA data). We find that only the pulsars born within the last 
$\sim 3\times 10^3$ yrs are able to provide fluxes of neutrons
which exceed the CR limit, provided that they accelerate nuclei 
with the efficiency $\xi = 1$ (see Eq.~\ref{eq2}).

\begin{figure}
  \vspace{6.truecm}
\includegraphics{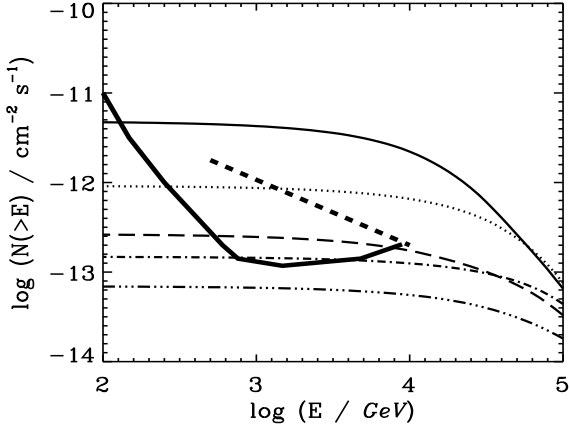}
  \caption{The integral spectra of $\gamma$-rays produced in hadronic
interactions of iron nuclei with the matter of the molecular cloud 
at the Galactic Center for the models considered in the text: I (full curve),
II (dotted), III (dot-dashed), and IV (dot-dot-dot-dashed), and V
(dashed). The thick-dashed and full curves show the sensitivity
of the operating HEGRA System and planned CANGAROO III, HESS, and
VERITAS systems for a 50-hour exposure to a single source (Lorenz~2001, and
references therein).}
      \label{fig3}
\end{figure}
\section{Discussion and Conclusion}

The SUGAR group estimates the flux of particles 
which causes reported excess of the cosmic ray particles in the energy range 
$10^{17.9}-10^{18.5}$ eV on $(9\pm
3)\times 10^{-14}$ m$^{-3}$ s$^{-1}$ (Bellido et al.~2001).
If this excess is caused by neutrons produced in the pulsar model discussed
here, then the expected flux of neutrons can be compared
with  the observed one. Basing on this normalization we predict the fluxes of
neutrinos and gamma-rays on Earth. This procedure allows us to derive the
free parameter of our model (i.e. the efficiency $\xi$ of iron acceleration
by the pulsar) and limit the age of the pulsar for other fixed 
parameters, $P, B, R_{\rm c}, n_{\rm c}, B_{\rm c}$, which are in fact
constrained by the observations. 
We consider five different sets of parameters describing our scenario:  
model (I) R = 10 pc, $n =
10^3$ cm$^{-3}$,  $B_{\rm c} = 10^{-4}$ G, $t_{\rm obs} = 10^4$ yr; (II) 
$t_{\rm obs} = 3\times 10^3$ yr and other  parameters as above; 
(III) $t_{\rm obs} = 10^3$ yr and other parameters as above;
(IV) R = 50 pc, $n = 10^2$ cm$^{-3}$, $B_{\rm c} = 3\times 10^{-5}$ G, 
$t_{\rm obs} = 10^4$ yr; and (V) $t_{\rm obs} = 3\times 10^3$ yr and
other parameters as in (IV). They all concern two sets of parameters for the
medium in which the pulsar is formed, and differ in the pulsar's age 
which is not constrained by any observations.
In all these models we assume that the
pulsar is born with $B = 6\times 10^{12}$ G and  $P_{\rm o} = 2$ ms.
Normalizing the predicted neutron flux to the observed 
excess of CR particles we derive the value of the parameter $\xi$ which has 
to be $\xi\approx 1$ (model I), $0.18$ (II), $0.03$ (III), $0.3$ (IV), and
$0.09$ (V).

\begin{table}
\caption{Gamma-rays and neutrinos from the Galactic Center.}     
\begin{tabular}{|c|c|c|c|c|}
\hline 
Model  & $N_\gamma$($>1$ TeV)& $N_\gamma$($>10$ TeV) 
& $N_\nu^{\rm a}$ & $N_\nu^{\rm na}$ \\
\hline      
 (I)  & $4.3\times 10^{-12}$ & $2.2\times 10^{-12}$  & 23 & 30 \\
\hline
(II)  & $8.7\times 10^{-13}$ & $6.6\times 10^{-13}$  & 11 & 16 \\
\hline
(III) & $1.4\times 10^{-13}$ & $1.25\times 10^{-13}$ & 4.2 & 7.1 \\
\hline
(IV)  & $2.5\times 10^{-13}$ & $1.7\times 10^{-13}$  & 5.3 & 8.8 \\
\hline
(V)  & $6.7\times 10^{-14}$ & $5.5\times 10^{-14}$  & 2.0 & 3.8 \\
\hline
\end{tabular}
\label{table1}
\end{table}

Using the above estimates for $\xi$ we can now predict the expected
fluxes of $\gamma$-rays and muon neutrinos and antineutrinos in 
the case of every
model. The  integral spectra of $\gamma$-rays from the GC region are 
presented in
Fig.~\ref{fig3}, together with the sensitivities of the present HEGRA
telescope system and the planned next generation of telescopes, i.e. CANGAROO
III, HESS, VERITAS. We also report in Table~\ref{table1} the $\gamma$-ray
fluxes above 1 TeV and 10 TeV in units cm$^{-2}$ s$^{-1}$.
Although the $\gamma$-ray spectra have a maximum above 10 TeV for all models, 
the $\gamma$-ray fluxes in the energy range  1-10 TeV produced in models, (I)
$\sim 2.1\times 10^{-12}$  cm$^{-2}$ s$^{-1}$, and (II) $\sim 2.1\times
10^{-13}$ cm$^{-2}$ s$^{-1}$, and probably also in (IV) $\sim 8\times 10^{-14}$
cm$^{-2}$ s$^{-1}$, should be observed by the future systems of Cherenkov 
telescopes of the CANGAROO III, HESS, and VERITAS type.  
Models (III) and (V) predict fluxes below the sensitivity limit
of these Observatories. Only the HEGRA Collaboration observed the Galactic 
disk including the GC region (P\"ohlhofer et al.~1999). The upper
limit on the possible sources, equals 1/4 Crab in the Galactic plane 
which is above the $\gamma$-ray flux
predicted even by the model (I). However, since the GC region can be observed
by this experiment only at zenith angles larger than $60^{\rm o}$, this limit
does not refer to the GC region.  

\begin{figure}
  \vspace{6.truecm}
\includegraphics{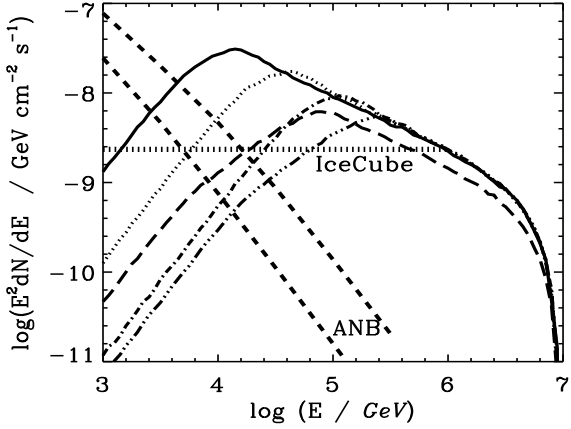}
  \caption{The differential spectra of muon neutrinos and antineutrinos
produced in hadronic interactions of iron nuclei with the matter of a
molecular cloud  at the Galactic Center for the models discussed in Fig.~3. 
The dashed curves indicate the atmospheric neutrino background, ANB,
(Lipari~1993) within a $1^{\rm o}$ of the source and the dotted line shows
the 3 yr sensitivity of the IceCube detector (Hill~2001).}   
\label{fig4}
\end{figure}

In Fig.~\ref{fig4} we show the muon neutrino and anti-neutrino spectra 
expected in the discussed model. At energies $> 10$ TeV, these spectra are
above the expected flux of atmospheric neutrino background (ANB) and also
above  the 3 yr sensitivity limit of the planned large size neutrino 
detector
IceCube (Hill~2001). We estimate the number of muon neutrino events during 
one year in the IceCube detector basing on the calculations of the 
likelihood of detecting such neutrinos by a detector with a 
surface area of 1 km$^2$ obtained by Gaisser \& Grillo~(1987). 
The results of these calculations are shown in Table~1. 
We distinguish the case of neutrinos coming to the neutrino detector from
directions close to the horizon, i.e. not absorbed by the Earth 
($N_{\nu}^{\rm na}$), and  neutrinos which arrive moving upward from the nadir
direction, i.e. partially absorbed by the Earth ($N_{\nu}^{\rm a}$) (for
absorption coefficients see Gandhi~2000). From Table~1 it is clear that 
the IceCube detector should detect a few up to several neutrinos per year from
the Galactic Center region provided that the excess of cosmic rays at $\sim
10^{18}$ eV from the GC region is caused by neutrons from desintegrations of
iron nuclei, accelerated by a very fast pulsar. The detection of the predicted
fluxes of neutrinos from the Galactic Center (or lack) will also place 
constraints on the recent model of extremely high energy cosmic ray
production in the pulsar scenario (Blasi, Epstein \& Olinto~2000), since 
the parent iron nuclei which inject neutrons with energies
$\sim 10^{18}$ eV, have to be accelerated to energies $\sim 10^{20}$ eV.

The EGRET detector on board the Compton GRO has detected a strong
$\gamma$-ray source with luminosity of $\sim 2\times 10^{37}$ erg s$^{-1}$ 
(Mayer-Hasselwander et al.~1998).
This emission seems to come from the extended region with the radius of 
$\sim 80$ pc around the GC. In another paper (in preparation) we
suggest that this emission can be 
explained in terms of the general scenario discussed here if 
a part of energy of the accelerated iron nuclei is transferred to the
relativistic positrons due to the resonant scattering in the pulsar shock
region (see Gallant \& Arons~1994). These positrons, accumulated in the 
cloud, produce high energy radiation in synchrotron and inverse Compton 
processes.

\section*{Acknowledgements}
This work is supported by the Polish KBN grant No. 5P03D02521.

\end{document}